\documentstyle[twoside,fleqn,espcrc2]{article}

\newcommand{\AmS}{{\protect\the\textfont2
  A\kern-.1667em\lower.5ex\hbox{M}\kern-.125emS}}

\title{Gauged Knizhnik-Zamolodchikov Equation}

\author{Ian I. Kogan, Alex Lewis\address{Department of Physics, 
University of Oxford\\ 
1 Keble Road, Oxford, OX1 3NP, United Kingdom}
and Oleg A. Soloviev\address{Physics Department, Queen Mary College 
\\
        Mile End Road, London, E1 4NS, United Kingdom}%
        } 

\begin{document}

\begin{abstract}
Correlation functions of gauged WZNW models are shown to satisfy a 
differential equation, which is a gauge generalization of the 
Knizhnik-Zamolodchikov equation.\end{abstract}

\maketitle

\section{INTRODUCTION}

A large class of 2D conformal field theories is described by 
gauged Wess-Zumino-Novikov-Witten models with the following action 
\cite{Gawedzki} 
\begin{eqnarray}
S(g, A)=S_{WZNW}(g)~+~{k\over2\pi}\int d^2z\mbox{Tr}[Ag^{-
1}\bar\partial g\nonumber\end{eqnarray}
\begin{equation}
-\bar A\partial gg^{-1}+Ag^{-1}\bar Ag-A\bar 
A],\end{equation}
where
\begin{eqnarray}
&&S_{WZNW}(g)={k\over8\pi}\int d^2z\mbox{Tr}g^{-1}\partial^{\mu}gg^{-
1}\partial_{\mu}g\nonumber\\ 
&&+
{ik\over12\pi}\int d^3z\mbox{Tr}g^{-1}dg\wedge g^{-1}dg\wedge g^{-
1}dg
\end{eqnarray}
and $g\in G$, $A,~\bar A$ are the gauge fields taking values in the 
algebra $\cal H$ of the 
diagonal group of the direct product $H\times H,~H\in G$. From now on 
we shall be dealing 
with the orthogonal decomposition of the Lie algebra $\cal G$ as
\begin{equation}
{\cal G}={\cal H}\oplus{\cal M},~~~~[{\cal H},{\cal H}]\subset{\cal 
H},~~~~
[{\cal H},{\cal M}]\subset{\cal M}.\end{equation}

There are different methods of studying correlation functions of gauged 
WZNW models such as 
the free field realization approach or the fermionization technik 
\cite{Naculich}. We shall 
pursue a different idea which is paralell to the analysis of the 
gravitational dressing of 2D 
field theories \cite{Klebanov}.

Our starting point are the equations of motion of the gauged WZNW 
model:
\begin{eqnarray}
\bar\nabla(\nabla gg^{-1})&=&0,\nonumber\\
\nabla gg^{-1}|_{\cal H}&=&0,\\
\bar\partial A~-~\partial\bar A~+~[A,\bar 
A]&=&0,\nonumber\end{eqnarray}
where
\begin{equation}
\bar\nabla=\bar\partial~+~\bar A,~~~\nabla=\partial~+~A.\end{equation}
Under the gauge symmetry, the WZNW primary fields $\Phi_i$ and the 
gauge fields $A,~\bar A$ 
transform respectively as follows
\begin{eqnarray}
\delta\Phi_i&=&\epsilon^a(t^a_i+\bar t^a_i)\Phi_i,\nonumber\\
\delta A&=&-\partial\epsilon-[\epsilon,A],\\
\delta\bar A&=&-\bar\partial\epsilon-[\epsilon,\bar 
A],\nonumber\end{eqnarray}
where $t^a_i\in{\cal H}$.

In order to fix the gauge invariance, we impose the following condition
\begin{equation}
\bar A=0.\end{equation}
Alternatively, one can choose any other gauge condition, say
\begin{equation}
A=0.\end{equation}
The physical sector of the quantum theory must not depend on the gauge 
choice. The gauge fixing (7) gives rise to the corresponding Faddeev-
Popov ghosts with the action
\begin{equation}
S_{ghost}=\int d^2x\mbox{Tr}(b\partial c).\end{equation}

In the gauge (7), the equations of motion take the following form
\begin{eqnarray}
\bar\partial J&=&0,\nonumber\\
\bar\partial A&=&0,\\ 
J|_{\cal H}&=&0,\nonumber\end{eqnarray}
where
\begin{equation}
J=-{k\over2}\partial gg^{-1}~-~{k\over2}gAg^{-1}.\end{equation}
Thus, $J$ is a holomorphic current in the gauge (7). Moreover, it has 
canonical commutation relations with the field $g$ and itself:
\begin{equation}
\{J^a(w),g(z)\}=t^ag(z)\delta(w,z),\end{equation}
\begin{eqnarray}
\{J^a(w),J^b(z)\}=f^{abc}J^c(z)\delta(w,z)~+~{k\over2}\delta^{ab}
\delta'(w,z).\nonumber\end{eqnarray}
The given commutators follow from the symplectic structure of the 
gauged WZNW model in the gauge (7). In this gauge, the field $A$ plays 
a role of the parameter $v_0$ of the orbit of the affine group $\hat G$ 
\cite{Alekseev}. Therefore, the symplectic structure of the original 
model in the gauge (7) follows from the symplectic structure of the 
original WZNW model \cite{Faddeev}.

The crucial point is that there are residual symmetries which survive 
the gauge fixing (7). Under these symmetries the fields $\Phi_i$ and 
the remaining field $A$ transform according to
\begin{eqnarray}
\tilde\delta\Phi_i&=&(\epsilon^A_Lt^A_i~+~\epsilon^a_R\bar 
t^a)\Phi_i,\nonumber\\ & & \\
\tilde\delta A&=&-\partial\epsilon_R~-
~[\epsilon_R,A],\nonumber\end{eqnarray}
where the parameters $\epsilon_L$ and $\epsilon_R$ are arbitrary 
holomorphic functions,
\begin{equation}
\bar\partial\epsilon_{L,R}=0.\end{equation}
In eqs.(13)the generators $t^A$ act on the left index of $\Phi_i$, 
whereas $\bar t^a$ act on the right index of $\Phi_i$. One can notice 
that the left residual group is extended to the whole group $G$, 
whereas the right residual group is still the subgroup $H$.

\section{WARD IDENTITIES}

Let us define dressed correlation functions
\begin{eqnarray}
&&\langle\langle\cdot\cdot\cdot\rangle\rangle\equiv\int 
{\cal D}\bar A{\cal D}A\langle\cdot\cdot\cdot\rangle\exp[-
{k\over2\pi}\int d^2z\nonumber\\ & & \\
&&\times\mbox{Tr}\{\bar Ag^{-1}\partial g+A\bar\partial gg^{-1}+Ag\bar 
Ag^{-1}+A\bar A\}],\nonumber\end{eqnarray}
where$\langle\cdot\cdot\cdot\rangle$ is the correlation function before 
gauging, which is found as a solution to the Knizhnik-Zamolodchikov 
equation.

In the gauge (7), the dressed correlation functions can be presented as 
follows
\begin{eqnarray}
&&\langle\langle\Phi_1(z_1,\bar z_1)\Phi_2(z_2,\bar 
z_2)\cdot\cdot\cdot\Phi_N(z_N,\bar z_N)\rangle\rangle\nonumber\\
&&=\int{\cal D}b{\cal D}c\exp(-S_{ghost})\nonumber\\
&&\times\int{\cal D}A\exp[-S_{eff}(A)]\\
&&\times\int{\cal D}g~\Phi_1(z_1,\bar z_1)\Phi_2(z_2,\bar 
z_2)\cdot\cdot\cdot\Phi_N(z_N,\bar z_N)\nonumber\\
&&\times\exp[-\Gamma(g,A)],\nonumber\end{eqnarray}
where $S_{eff}(A)$ is the effective action of the field $A$ and 
$\Gamma(g,A)$ is formally identical to the original gauged WZNW action 
in the gauge (7). The action $S_{eff}$ is non-local and can be obtained 
by integration of the following variation (which follows from the Wess-
Zumino anomaly condition)
\begin{equation}
\partial{\delta S_{eff}\over\delta A^a}+f^{abc}A^c{\delta 
S_{eff}\over\delta A^b}=\tau\bar\partial A^a.\end{equation}
Here the constant $\tau$ is to be defined from the consistency 
condition of the gauge (7), which is
\begin{equation}
J_{tot}\equiv{\delta Z\over\delta\bar A^a}=0, ~~~~a=1,2,...,\dim H,
\end{equation}
at $\bar A=0$. Here $Z$ is the partition function of the gauged WZNW 
model. Condition (18) amounts to the vanishing of the central charge of 
the affine current $J_{tot}$ which is a quantum analog of $J|_{\cal 
H}$. This in turn means that $J|_{\cal H}$ is a first class constraint 
\cite{Schnitzer}. In order to use this constraint, we need to know the 
OPE of $A$ with itself. This can be derived as follows. Let us consider 
the identity
\begin{eqnarray}
&&\tau\langle\langle\bar\partial A(z)A(z_1)\cdot\cdot\cdot 
A(z_N)\rangle\rangle
=\int{\cal D}A
A(z_1)\cdot\cdot\cdot A(z_N)\nonumber\end{eqnarray}
\begin{equation} 
\times\left[\partial{\delta S_{eff}\over\delta 
A^a(z)}+f^{abc}A^c{\delta S_{eff}\over\delta A^b(z)}\right]
\mbox{e}^{-S_{eff}}.\end{equation}
Here we used relation (17). Integrating by parts in the path integral, 
we arrive at the following formula
\begin{eqnarray}
&&\tau\langle\langle A^a(z)A^{a_1}(z_1)\cdot\cdot\cdot 
A^{a_N}(z_N)\rangle\rangle\nonumber\\ 
&&={1\over2\pi i}\sum^N_{k=1}[{-\delta^{aa_k}\over(z-z_k)^2} 
\langle\langle A^{a_1}(z_1)\cdot\cdot\cdot\hat 
A^{a_k}(z_k)\cdot\cdot\cdot A^{a_N}(z_N)\rangle\rangle \nonumber 
\end{eqnarray}
\begin{equation}
+{f^{aa_kb}\over z-z_k}\langle\langle A^{a_1}(z_1)\cdot\cdot\cdot 
A^{a_k}(z_k)\cdot\cdot\cdot A^{a_N}(z_N)\rangle\rangle],\end{equation}
where $\hat A(z_k)$ means that the field $A(z_k)$ is removed from the 
correlator. In the derivation of the last equation we used the 
following identity
\begin{equation}
\bar\partial{1\over z-z_k}=2\pi i\delta^{(2)}(z-z_k).\end{equation}

From eq. (20), it follows that
\begin{equation}
\tau A^a(z)A^b(0)={1\over2\pi i}\left[-{\delta^{ab}\over 
z^2}+{f^{abc}\over z}A^c(0)\right]+\mbox{reg}.\end{equation}
Along with condition (18), the equation (22)gives the expression for 
$\tau$
\begin{equation}
\tau={i(k+2c_V(H))\over4\pi}.\end{equation}

We proceed to derive the Ward identity associated with the residual 
symmetry (13). The Ward identity comes along from the variation of eq. 
(16) under transformations (13). Because it must be zero, we obtain the 
following relation
\begin{eqnarray}
\sum^N_{k=1}\bar t^a_k\delta(z,z_k)\langle\langle\Phi_1(z_1,\bar 
z_1)\Phi_2(z_2,\bar z_2)\cdot\cdot\cdot\Phi_N(z_N,\bar 
z_N)\rangle\rangle\nonumber\end{eqnarray}
\begin{equation}
\end{equation}
\begin{eqnarray}
+\tau\langle\langle\bar\partial_{\bar z}A^a(z)\Phi_1(z_1,\bar 
z_1)\Phi_2(z_2,\bar z_2)\cdot\cdot\cdot\Phi_N(z_N,\bar 
z_N)\rangle\rangle=0.\nonumber\end{eqnarray}
This yields
\begin{eqnarray}
2\pi\tau\langle\langle A^a(z)\Phi_1(z_1,\bar z_1)\Phi_2(z_2,\bar 
z_2)\cdot\cdot\cdot\Phi_N(z_N,\bar 
z_N)\rangle\rangle\nonumber\end{eqnarray}
\begin{equation}
\end{equation}
\begin{eqnarray}
=i\sum^N_{k=1}{\bar t^a_k\over z-z_k}\langle\langle\Phi_1(z_1,\bar 
z_1)\Phi_2(z_2,\bar z_2)\cdot\cdot\cdot\Phi_N(z_N,\bar 
z_N)\rangle\rangle,\nonumber\end{eqnarray}
which in turn gives rise to the OPE between the gauge field $A^a$ and 
$\Phi_i$
\begin{equation}
A^a(z)\Phi_i(0)={2\over k+2c_V(H)}{\bar t^a_i\over z}\Phi_i(0)   
.\end{equation}

Now we are in a position to define the product $A(z)\Phi_i(z,\bar z)$. 
Indeed, we can define this according to the following rule
\begin{equation}
A^a(z)\Phi_i(z,\bar z)=\oint{d\zeta\over2\pi i}{A^a(\zeta)\Phi_i(z,\bar 
z)\over\zeta-z},\end{equation}
where the nominator is understood as OPE (26). Formula (27) is a 
definition of normal ordering for the rpoduct of two operators.

\section{GAUGE DRESSING}

Let us come back to equation (11). At the quantum level it can be 
presented in the following form
\begin{equation}
\partial g+\eta gA+(2/\kappa)Jg=0.\end{equation}
Here $\eta$ and $\kappa$ are some renormalization constants due to 
regularization of singular products $gA$ and $Jg$.

Variation of (28) under the residual symmetry with the parameter 
$\epsilon_R$ gives rise to the following relation
\begin{equation}
\left[1-\eta\left(1-{c_V(H)\over 
k+2c_V(H)}\right)\right]\partial\epsilon_R(z)g(z)=0.\end{equation}
From this relation we find the renormalization constant $\eta$
\begin{equation}
\eta={k+2c_V(H)\over k+c_V(H)}.\end{equation}
In the classical limit $k\to\infty$, $\eta\to1$.

In the same fashion, we can compute variation of (28) with respect to 
the residual symmetry with the parameter $\epsilon_L$. This fixes the 
constant $\kappa$ as follows
\begin{equation}
\kappa={1\over k+c_V(G)}.\end{equation}
Note that the given expression for $\kappa$ is consistent with the 
condition that the combination $\partial+\eta A$ acted on $g$ as a 
Virasoro generator $L_{-1}$:
\begin{equation}
L_{-1}g={2J^A_{-1}J^A_0\over k+c_V(G)}g.\end{equation}

All in all, with the regularization given by eq. (27) and the Ward 
identity (25) the equation (28) gives rise to the following 
differential equation
\begin{eqnarray}
\left\{{1\over2}{\partial\over\partial z_i}+\sum^N_{j\ne 
i}\left[{t^A_it^A_j\over k+c_V(G)}-{\bar t^a_i\bar t^a_j\over 
k+c_V(H)}\right]{1\over z_i-z_j}\right\}\nonumber\end{eqnarray}
\begin{equation} 
\times\langle\langle\Phi_1(z_1,\bar z_1)\Phi_2(z_2,\bar 
z_2)\cdot\cdot\cdot\Phi_N(z_N,\bar z_N)\rangle\rangle=0,\end{equation}
where $t^A_i\in{\cal G}$ and $\bar t^a_i\in{\cal H}$.

Note that one could have started with the gauge (8). In this case one 
would have derived a similar equation for the antiholomorphic 
coordinate $\bar z$:
\begin{eqnarray}
\left\{{1\over2}{\partial\over\partial\bar z_i}+\sum^N_{j\ne 
i}\left[{\bar t^A_i\bar t^A_j\over k+c_V(G)}-{ t^a_i t^a_j\over 
k+c_V(H)}\right]{1\over \bar z_i-
\bar z_j}\right\}\nonumber\end{eqnarray}
\begin{equation} 
\times\langle\langle\Phi_1(z_1,\bar z_1)\Phi_2(z_2,\bar 
z_2)\cdot\cdot\cdot\Phi_N(z_N,\bar z_N)\rangle\rangle=0.\end{equation}
Because correlation functions do not depend on the gauge, we arrive at 
the conclusion that $\langle\langle\Phi_1(z_1,\bar z_1)\Phi_2(z_2,\bar 
z_2)\cdot\cdot\cdot\Phi_N(z_N,\bar z_N)\rangle\rangle$ obey both 
equation (33) and (34).

Equations (33) and (34) are our main result. By solving them, one can 
find dressed correlation functions in the gauged WZNW model. The 
solution can be expressed as products of the correlation functions in 
the WZNW model for the group $G$ at level $k$ and $H$ atlevel $-k-
2c_V(H)$. In particular, for the two-point function the equation yields
\begin{equation}
\partial\langle\langle\Phi_i(z,\bar 
z)\Phi_j(0)\rangle\rangle\end{equation}
\begin{eqnarray}
=-2\left[{t^A_it^A_j\over k+c_V(G)}-{\bar t^a_i\bar t^a_j\over 
k+c_V(H)}\right]{1\over z}\langle\langle\Phi_i(z,\bar 
z)\Phi_j(0)\rangle\rangle\nonumber.\end{eqnarray}
By the projective symmetry, the two-point function has the following 
form
\begin{equation}
\langle\langle\Phi_i(z,\bar 
z)\Phi_j(0)\rangle\rangle={G_{ij}\over|z|^{4\Delta_i}},\end{equation}
where $\Delta_i$ is the anomalous conformal dimension of $\Phi_i$ after 
the gauge dressing and $G_{ij}$ is the Zamolodchikov metric which can 
be diagonalized. After substitution of expression (36) into eq. (35), 
and using the fact that, as a consequence of the residual symmetry 
(13), the dressed correlation functions must be singlets of both the 
left residual group $G$ and the right residual group $H$, we find
\begin{equation}
\Delta_i={c_i(G)\over k+c_V(G)}-{c_i(H)\over k+c_V(H)},\end{equation}
where $c_i(G)=t^A_it^A_i,~c_i(H)=\bar t^a_i\bar t^a_i$.

Eqs. (33) and (34) describe the gauge dressing of correlation functions 
of primary operators. In an ordinary WZNW model, correlators of primary 
operators contain complete information about the system as correlators 
of any descendant operators can be expressed in terms of correlation 
functions of primary operators alone \cite{Knizhnik}. The point to be 
made is that descendants of the WZNW model may become primaries of the 
gauged theory. In the case of a gauged WZNW model, the Ward identities 
of the ungauged theory are no longer applicable. Nevertheless, dressed 
correlation functions of descendant operators can still be represented 
in terms of dressed correlators of primary operators. This is due to 
the equation (28), which allows us to express $\partial gg^{-1}$ as a 
combination of the affine current $J$ and the product $A^{\bar 
a}\phi^{a\bar a}$, where $\phi^{a\bar a}=\mbox{Tr}t^ag^{-1}t^{\bar 
a}g$. Since we know that there are the Ward identities for $J$ and $A$, 
we can use them to derive a Ward identity for $\partial gg^{-1}$, which 
will relate dressed correlation functions of descendants with 
correlators of primaries. 

\section{CONCLUSION}

For the gauged WZNW models, which form a particular class of 2D gauged 
theories, we have derived differential equations which, in principle, 
allow one to find correlation functions of these models. In 
\cite{Kogan} we have used our equations to show that the $SL(2)/U(1)$ 
conformal blocks can be expressed as products of $SL(2)$ and $U(1)$ 
conformal blocks. The detailed analysis of the dressed four-point 
function in the $SL(2)/U(1)$ coset construction can also be found in 
\cite{Kogan}.

\end{document}